\documentclass[12pt]{iopart}
\usepackage{epsfig,epsf,graphics}
\usepackage{graphicx}
\def\e{\varepsilon}

\def\prl{{\it Phys. Rev. Lett.}\ }

\def\be{\begin{equation}}
\def\ee{\end{equation}}
\def\bea{\begin{eqnarray}}
\def\eea{\end{eqnarray}}

\begin{document}
\title{The theory of the "0.7 anomaly" in quantum point contacts}
\author{Yigal Meir}
\address{Physics Department, Ben Gurion University, Beer Sheva 84105, ISRAEL}
%
\bigskip
\bigskip
\bigskip
\begin{abstract}
The phenomenology of the "0.7 anomaly" in quantum point contacts
is fully explained in terms of a quasi-localized state, which
forms as the point contact opens up. Detailed numerical
calculations within spin-density functional theory indeed confirm
the emergence of such a state. Quantitative calculations of the
conductance and the noise are obtained using a model based on these
observations, and are in excellent agreement with existing
experimental observations.
\end{abstract}
%
\section{Introduction and summary}
The "0.7 anomaly", the subject of this collection, has been a long
standing puzzle since it was first realized \cite{thomas}
 that this is a generic
phenomenon, which in fact had been already evident in the first
experiments on conductance quantization in quantum points contacts
(QPCs)\cite{vanWees,Wharam}. In this article I will demonstrate
that existing experimental observations are consistent with a
simple model, based on the formation of a quasi-bound state in the
QPC as it opens up. I will first discuss the model and how it
explains the "0.7 anomaly" rather briefly, and then in later sections
will elaborate on a first principle calculation that indeed gives
rise to the formation of such a state (section II) and on
quantitative calculations of the conductance and the noise using a
model based on these observations (section III).

The main assumption (to be substantiated in the next section)
underlying the model is the emergence of a quasi-localized state
at the QPC near pinch-off. With this assumption the following
physics emerges:
\begin{itemize}
    \item The conductance is reduced from its universal value
    of $G_0\equiv 2e^2/h$, basically due to Coulomb blockade. As an electron
    is transferred through the quasi-localized state, it reduces the
    probability that another electron, with opposite spin, will be
    simultaneously transferred. Consequently, depending on parameters,
    the conductance will have a constant value between $0.5 G_0$ and $G_0$
    for any value of gate voltage in the Coulomb blockade regime.
    \item  As the gate voltage is further increased the Coulomb
    blockade energy is overcome and the conductance reaches the
    value $G_0$. As we will see in the next section, the localized
    state disappears for such gate voltages.
    \item   As magnetic field is increased the quasi-localized state
    energy splits. Similar to the physics of quantum dots, the
    conductance will decrease to $0.5 G_0$ and the separation in
    gate voltage between the "0.7" step and the first plateau will
    increase linearly.
    \item  The conductance around the "0.7" plateau may be
    thought of as carried by two channels, one with almost perfect
    transmission and one with reduced transmission. This leads
    to the reduction of the shot noise in this regime, compared to
    the situation where both channels carry the same conductance.
    \item  At "high" temperatures the spin of the quasi-localized state
    is fluctuating among all degenerate directions. As temperature
    is reduced below the Kondo temperature, at zero magnetic field,
    this local spin will be screened  by the lead electrons -- the
    Kondo effect -- enhancing the conductance beyond
    its high temperature value towards the unitarity limit, $G_0$.
    A small magnetic field, such as the Zeeman splitting is of the
    order of the Kondo temperature, will polarize the spin.
    \item  Formation of a quasi-localized state will also occur at large magnetic
    fields, where the lower spin branch of the second QPC mode
    crossed the higher spin branch of the first model, giving rise,
    in this case, to a "1.2" plateau. Due to the finite magnetic field,
    Kondo physics will not be relevant here at low temperatures.
    \item Sometimes a localized state is also formed at the
    opening of the second subband, giving rise to a "1.7" plateau.
    Unlike the generic situation in the first subband, we find the
    formation of this state in the second subband to be sensitive
    to parameters.
\end{itemize}
   All these observations are in agreement with experiments.
\section{Formation of a magnetic impurity - a density functional calculation}
In this spin-density functional calculation (SDFT)\cite{rejec},
which generalizes an earlier SDFT calculation \cite{hirose}, we
treat the two dimensional electron gas (2DEG), the electrodes and
the donor layer as a set of three electrostatically coupled
 two-dimensional systems (Fig.~1a). We assume the donor
layer is uniform and fully ionized, and the gates are kept in
some voltage (gate-voltage) with respect to the 2DEG. Then,
according to spin-density-functional theory (SDFT)\cite{HK}, the
properties of the system can be uniquely determined in terms of
the spin densities of the 2DEG and of the distribution of charge
on the electrodes. We treat the 2DEG quantum-mechanically by
including its kinetic energy and exchange-correlation energy  in
the energy functional, taking into account the GaAs effective mass
and dielectric constant. We used the local spin-density
approximation for the exchange-correlation functional, as
parameterized in Ref.\cite{tanatar} (For more details see
Ref.\cite{rejec}). The effective self-consistent potential
experienced by electrons in the 2DEG is depicted in Fig.~1(b).

\begin{figure}[ht]
\begin{center}
\includegraphics [bb=-8 61 559 829,width=1\textwidth,angle=0]{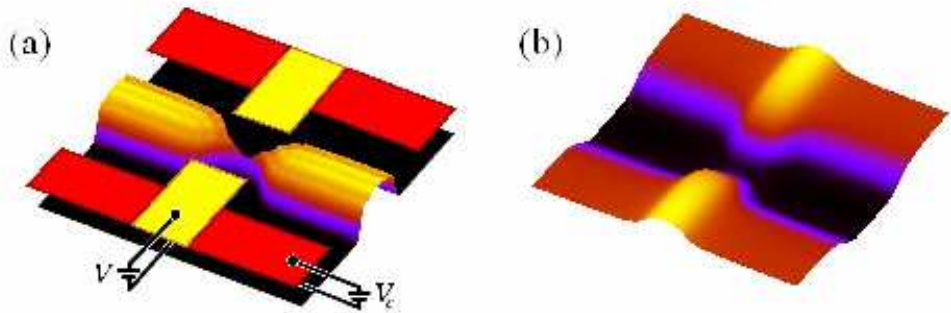}
\vskip -14.5 truecm \caption {\label{fig1} The gate electrodes for the SDFT calculation (a). 
The red gates define the maxima possible number of modes, while the yellow gate define the
QPC. An example of electron density in the wire is also depicted. (b) The effective potential
in the 2DEG.}
\end{center}
\end{figure}

We want to be able to include the possibility of a quasi-localized
state in the calculations. If such a state indeed forms, it will
necessarily be occupied by a single spin. To be able to
incorporate such a solution within SDFT, we allow solutions of the
Kohn-Sham equation \cite{Kohn} which break spin-symmetry. Indeed
the lowest energy solution, as the QPC opens up, is a
spin-polarized state (though the spin-direction is arbitrary) -- As
the effective QPC barrier is lowered the two semi-infinite
electrons gases on its two sides start to overlap each other and
the density on top of the QPC increases. Once this density is
enough to support a full electron, the lowest energy solution
describes a quasi-localized electron on top of the QPC (Fig.~2).
This solution was found to be the ground state for all the range
of parameters we have checked, thus supporting the idea that the
formation of such a magnetic impurity in the opening of the first
mode is a generic effect.

\begin{figure}[ht]
\begin{center}
\includegraphics [bb=192 -359 799 429,width=0.6\textwidth,angle=0]{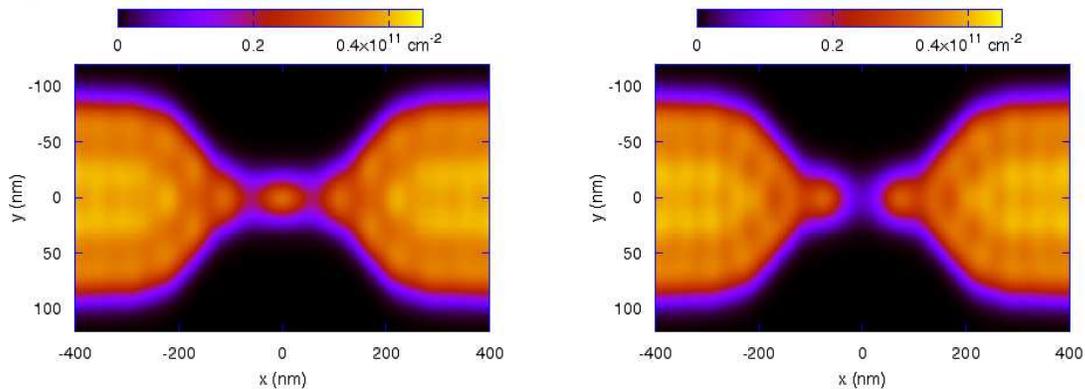}
\vskip -5 truecm \caption {\label{fig2} The spin up (a) and spin down (b) density distribution
near the gate voltage corresponding to the 0.7 anomaly, indicating formation of a magnetic
impurity on top of the QPC.}
\end{center}
\end{figure}

A similar solution is found sometimes, but not always, at the
opening of the second mode (as electrons occupying the lowest mode
screen out the Coulomb interactions), and the energy gained by the
formation of the quasi-localized state is significantly lower,
suggesting that a feature in the second plateau may be present,
but sensitive to details. On the other hand we also find a generic
localized state solution in large Zeeman fields, where the higher
spin state in the lower mode and lower spin state in the next mode
become degenerate (see Fig.~3). In this case we expect reduction
of the conductance from its $1.5G_0$ value, but no
spin-degeneracy. This is consistent with the observation of the
"analog" states \cite{analog}.

\begin{figure}[ht]
\begin{center}
\includegraphics [bb=192 -409 799 379,width=0.6\textwidth,angle=0]{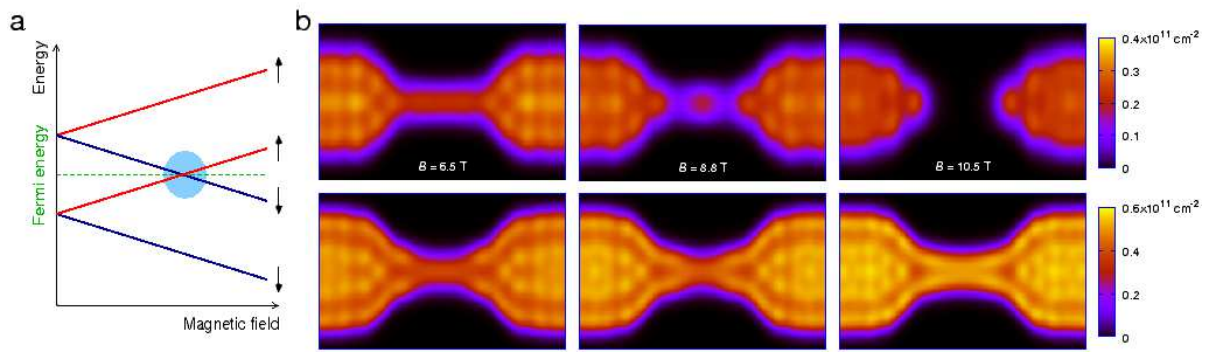}
\vskip -6 truecm \caption {\label{fig3} Formation of a quasi-localized state at 
the crossing of two spin sub-bands at large magnetic field. The two spin densities (top
and bottom) are shown at three values of magnetic field, demonstrating the formation
of that state only at a special value of the field.}
\end{center}
\end{figure}

While at high temperatures the localized electron fluctuates
between all possible spin-directions, so that the state is
instantaneously polarized, one expects that as the temperature is
lowered below the Kondo temperature, this spin will be screened by
the electrons in the leads. Unfortunately, such a state is beyond
the capability of SDFT in its local approximation (similar to the
Hydrogen molecule problem). In addition, SDFT also cannot, in
principle, give the correct dynamical properties of the system, in
particular the conductance (see e.g. Ref.\cite{car}). In order to
evaluate transport properties and also include the Kondo effect,
we use, in the next section, the static properties of the system,
as obtained from SDFT, to write down an effective Hamiltonian from
which one may calculate these properties using standard many-body
techniques.

\section{The effective model and transport properties}
Our SDFT results indicate that even an initially smooth QPC
potential can produce a narrow quasi-bound state, resulting in a
spin bound at the center of the QPC. We \cite{meir02} thus model
the QPC and its leads by the generalized Anderson Hamiltonian
\cite{Anderson}

\begin{eqnarray}
 H &=& \sum_{\sigma;k{\in L,R} }\!\!\! \e_{k\sigma}
{\bf c}_{k\sigma}^{\dagger} {\bf c}_{k\sigma}
  + \sum_\sigma\! \e_{\sigma}
{\bf d}_{\sigma}^{\dagger} {\bf d}_{\sigma}
  +  U {\bf n}_\uparrow {\bf n}_\downarrow \nonumber\\
  &+& \!\!\!\sum_{\sigma;k{\in L,R} }\!\! [V_{k\sigma}^{(1)}
      (1 - {\bf n}_{\bar{\sigma}})
     {\bf c}_{k\sigma}^{\dagger} {\bf d}_{\sigma}
                            + V_{k\sigma}^{(2)}
       {\bf n}_{\bar{\sigma}}
     {\bf c}_{k\sigma}^{\dagger} {\bf d}_{\sigma}
     + H.c.]
\label{HA}
\end{eqnarray}
where ${\bf c}_{k\sigma}^{\dagger} ({\bf c}_{k\sigma})$ creates
(destroys) an electron with momentum $k$ and spin $\sigma$ in one
of the two leads $L$ and $R$, ${\bf d}_{\sigma}^{\dagger} ({\bf
d}_{\sigma})$ creates (destroys) a spin-$\sigma$ electron on ``the
site", {\it i.e.} the quasi-bound state at the center of the QPC,
and ${\bf n}_{\sigma} = {\bf d}_{\sigma}^{\dagger} {\bf
d}_{\sigma}$. The hybridization matrix elements,
$V_{k\sigma}^{(1)}$ for transitions between 0 and 1 electrons on
the site and $V_{k\sigma}^{(2)}$ for transitions between 1 and 2
electrons, are taken to be step-like functions of energy,
mimicking the exponentially increasing transparency 
(the position of the step defines our zero of energy). Physically,
we expect $V_{k\sigma}^{(2)} < V_{k\sigma}^{(1)}$, as the Coulomb
potential of an electron already occupying the QPC will reduce the
tunneling rate of a second electron through the bound state. In
the absence of magnetic field the two spin directions are
degenerate, $\e_\downarrow=\e_\uparrow=\e_0$.

To obtain a quantitative estimate of the conductance we note that
the relevant gate-voltage range corresponds to the singly occupied
state regime. We therefore perform a Schrieffer-Wolff
transformation\cite{Schrieffer} to obtain the Kondo Hamiltonian
\cite{Kondo}

\begin{eqnarray}
H&=& \!\!\sum_{k\sigma \in L,R}
\varepsilon_{k\sigma}c^{\dagger}_{k\sigma}c_{k\sigma}+\!\!\!\!\!\!\!
\sum_{k,k'\sigma \in L,R}
(J^{(1)}_{k\sigma;k'\sigma}-J^{(2)}_{k\sigma;k'\sigma})
c^{\dagger}_{k\sigma}c_{k'\sigma}
\nonumber\\
 & +&2 \!\!\!\sum_ {k,k'\sigma\sigma' \in L,R}
(J^{(1)}_{k\sigma;k'\sigma'}+J^{(2)}_{k\sigma;k'\sigma'})
c^{\dagger}_{k\sigma}\vec{\sigma}_{\sigma\sigma'}
c_{k'\sigma'}\cdot \vec{S} \nonumber
\end{eqnarray}
\begin{equation}\label{int}
    J^{(i)}_{k\sigma;k'\sigma'}=
\frac{(-1)^{(i+1)}}{4}[\frac{V^{(i)}_{k\sigma}
V^{(i)}_{k'\sigma'}}
 {\varepsilon_{k\sigma}-\varepsilon^{(i)}_{\sigma}}+
 \frac{V^{(i)}_{k\sigma}V^{(i)}_{k'\sigma'}}
 {\varepsilon_{k'\sigma'}-\varepsilon^{(i)}_{\sigma'}}]
\end{equation}
The potential scattering term (first line), usually ignored in
Kondo problems, is crucial here, as it gives rise to the large
background conductance at high temperature. The magnetic field B,
defining the $z$-direction, enters the problem via the Zeeman
term, $S_z B$. The couplings are assumed to be exponentially
increasing as the QPC is opened. (In the above and in the
following $B$ and $T$ denote the corresponding energies, $g\mu_B
B$ and $k_BT$, respectively, where $k_B$ is the Boltzman cosntant,
$\mu_B$ is the Bohr magneton, and with the appropriate
$g$-factor.) Since $V^{(1)}\gg V^{(2)}$ then also $J^{(1)}\gg
J^{(2)}$.

The main calculational problem is that $J^{(1)}$ is not small and
thus cannot be used as a small expansion variable. To overcome
this we note that the model can be solved exactly in the large $B$
limit, where the spin channels are decoupled. Thus we \cite{golub}
perform an expansion around large fields (the small parameter is
$\exp(-B/T)$), which merges with the perturbation expansion in
small $J$ for zero magnetic field. Interpolating between these
limits, the conductance can be written

\begin{eqnarray}
G&=&\frac{e^2}{h}(T_1 +T_2) \label{G}  \nonumber\\
T_i&=&\frac{\tilde{g_i^2}}{1+\tilde{g_i^2}}. \label{G}
\end{eqnarray}
Where
 \begin{equation} \tilde{g_i^2}\equiv g_i^2 + \frac{B}{T
\sinh\frac{B}{T}} \frac{(g_1+g_2)^2}{1+(g_1+g_2)^2}
\label{gtilde},
\end{equation}
and $g_i=4\pi\nu J^{(i)}$.

At low temperature the Kondo effect develops, which leads
to
\begin{equation}
\tilde{g_2^2} \rightarrow
\tilde{g_2^2}+g_2^2(\frac{1}{2}-\frac{B}{T \sinh\frac{B}{T}}) +
G_2^{RG}, \label{gi}
\end{equation}
with
\begin{equation}
G_2^{RG}=\frac{1}{(\ln\frac{\sqrt{B^2+T^2}}{T_K})^2}\frac{\pi^2}{8}
(1+\frac{2 B}{T \sinh\frac{B}{T}}) \label{G2RG}
\end{equation}
with the Kondo temperature $T_K\simeq U\exp(-\pi/g_2)$. The Kondo
contribution enhances the contribution of the second channel, and
gives rise to the merging of the "0.7" feature with the first
$2e^2/h$ conductance step. As pointed out in \cite{meir02}, the
resulting $T_K$ increases exponentially with $\e_F$, in agreement
with the experimental observation that $T_K$ increases
exponentially with the gate voltage \cite{cron}. The resulting
conducting is plotted in Fig.~4a, giving rise, as expected to
plateaus in the conductance around $G\simeq0.7G_0$, which increase
with decreasing temperature (due to the perturbative nature of
the calculation a spurious nonmonotonicity appears in the conductance).
 Since the latter is due to the Kondo
effect, a zero-bias anomaly in the nonlinear conductance will
develop at low temperatures, as seen experimentally
\cite{cron}.

\begin{figure}[ht]
\begin{center}
\includegraphics [bb=22 -9 649 779,width=0.65\textwidth,angle=-90]{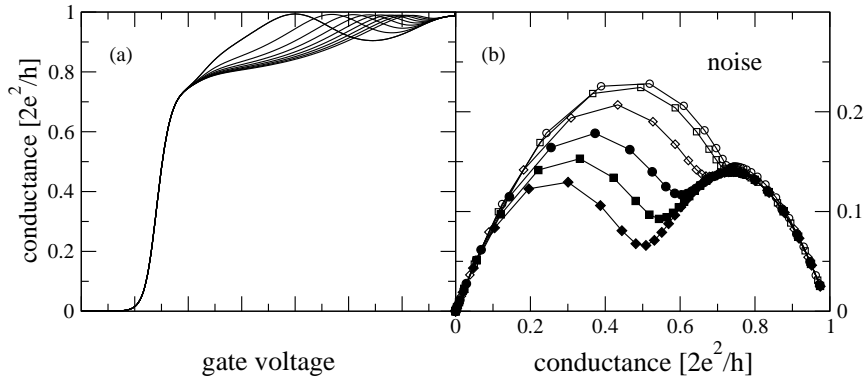}
\vskip -5 truecm \caption {\label{fig4} The conductance and the noise, resulting from
the approximate calculations presented here and in \cite{golub}.}
\end{center}
\end{figure}

Using the values of the transmission for the two channels,
Eq.\ref{G}, one can also evaluate the shot noise, which is
depicted in Fig~4b, giving rise to a dip in the noise around the
"0.7" anomaly, again consistent with experiments \cite{dicarlo}.

\section{Conclusions}
We have demonstrated that all the experimental data, as far as we
know, can be explained using a simple model that invokes the
localization of an electron in the QPC near pinch-off. The
emergence of such a quasi-localized state has been corroborated by
spin-density functional calculations.

\section*{Acknowledgement}
This work has been supported by the Bi-National US-Israel Science
Foundation.

\end{document}